\documentstyle[aps,amssymb]{revtex}
\newcommand{\be}{\begin{equation}}
\newcommand{\ee}{\end{equation}}
\newcommand{\ba}{\begin{eqnarray}}
\newcommand{\ea}{\end{eqnarray}}
\title{ Self-similar behavior of pre-turbulent fluctuations}
\author{R.~Collina$^{1,2}$, R.~Livi$^{3,4}$ and A.~Mazzino$^{5,1,6}$}
\address{
$^1$ Diartimento di Fisica, Universit\'a di Genova, Via Dodecaneso
33, I--16146 Genova, Italy\\
$^2$ Istituto Nazionale di Fisica Nucleare, Sez. di Genova, Via Dodecaneso
33, I--16146 Genova, Italy,\\
$^3$ Dipartimento di Fisica, Universit\'a di Firenze, Via Sansone 1,
I--50019 Firenze, Italy\\
$^4$ Istituto Nazionale di Fisica della Materia, UdR di Firenze e Istituto 
Nazionale di Fisica Nucleare, Sez. di Firenze, Via Sansone 1,
I--50019 Firenze, Italy\\
$^5$ CNR-ISAC, Strada Provinciale Lecce-Monteroni, I--73100 Lecce, Italy\\
$^6$ Istituto Nazionale di Fisica della Materia, UdR di Genova, Via Dodecaneso 33, I--16146 Genova, Italy\\}
\date{\today}            
\begin{document} 
\draft
\maketitle
\begin{abstract}
The random forced Navier-Stokes equation can be obtained as a variational
problem of a proper action. In virtue of incompressibility,
the integration over transverse components of the fields allows to cast
the action in the form of a large deviation functional.
Since the hydrodynamic operator is nonlinear, the functional integral
yielding the statistics of fluctuations can be practically computed by
linearizing around a physical solution of the
hydrodynamic equation. We show that this procedure yields
the dimensional scaling predicted by K41 theory at the lowest
perturbative order, where the perturbation parameter is the inverse
Reynolds number.  This result is valid over a finite
spatio-temporal domain, where the physical solution can be considered
as stationary.
\end{abstract}
\section{Introduction}
A field theoretic approach to the study of the random stirred Navier-Stokes
equation (NSE) can be traced back to the seminal paper by Martin, Siggia and
Rose \cite{MSR}. This was the starting point for the application of many
perturbative strategies, e.g. diagramatic expansions and renormalization
group methods \cite{ren}. The many technical difficulties encountered in
developing such approaches avoided to gather conclusive achievements.
For recent developments and applications the reader can be addressed to
\cite{Dev}. In this paper we show that one step
forward along this field-theoretic approach allows one to cast the
action associated with the NSE into the form
of a large deviation functional. Recently, large-deviation theory
has  scored sensible success in describing fluctuations in stationary
non-equilibrium regimes of various microscopic models \cite{BDGJL}.
This approach is mainly based on the extension of
the time-reversal conjugacy property introduced by Onsager and
Machlup \cite{OM} to stationary non-equilibrium states.
In practice, thermal fluctuations in irreversible stationary processes
can be traced back to a proper hydrodynamic description derived from
the microscopic evolution rules.
The action functional has the quadratic form
\be
I_{[(t_1,t_2)]}(\rho) = \frac{1}{2} \int_{t_1}^{t_2} dt\,\,
\langle W, K(\rho) \,\, W\rangle
\label{onsag}
\ee
where $\rho(t,\vec x)$ represents in general a vector of thermodynamic
variables depending on time $t$ and space variables $\vec x$.
The symbol $\langle \cdot \, ,\, \cdot \rangle$ denotes the integration
over space variables.
$W$ is a hydrodynamic evolution operator acting on
$\rho$: it vanishes when $\rho$ is equal to the stationary solution
$\bar\rho$, which is assumed to be unique. The positive kernel
$K(\rho)$ represents the stochasticity of the system
at macroscopic level.
According to the large deviation-theory the entropy $S$ of a stationary
non-equilibrium state is related to the action functional $I$ as
follows:
\be
S(\rho) = \inf_{\hat\rho} I_{[-\infty,0]} (\hat\rho)
\ee
where the minimum is taken over all trajectories connecting $\bar\rho$
to $\rho$.

For our purposes it is enough to consider that the action functional
$I$ provides a natural measure for statistical fluctuations in
non-equilibrium stationary states, so that, formally, any statistical
inference can be obtained from $I$.
Indeed, from the very beginning we have to deal with a hydrodynamic
formulation, namely the random stirred NSE: in the next Section
we will argue that an action functional of the form (\ref{onsag}) can be
obtained by field-theoretic analytic calculations.

In particular,
explicit integration over all longitudinal components of the velocity
field and over the associated auxiliary fields can be performed.
This allows to obtain a hydrodynamic evolution operator $W$ which depends
only on the transverse components of the velocity field $v_T(t,\vec x)$.
Moreover,
the positive kernel $K$ amounts to the inverse correlation function
of the stochastic source, while any dependence on the form of the
pressure tensor and of the noise does not enter
in the determination of $W$. Accordingly, many of 
the technical difficulties characterizing standard perturbative methods
and diagramatic expansions have been removed.

On the other hand, we have to face with new difficulties. 
The hydrodynamic operator appearing in the
large deviation functional is nonlinear, so that
functional integration is unfeasible. One should
identify a stationary solution $ {\bar v}_T(t,\vec x)$ 
of the associated hydrodynamic
equation and linearize the hydrodynamic operator around
such a solution. Then, functional integration could be performed
explicitly on the ``fluctuation'' field.
In order to be well defined this
approximate procedure would demand the uniqueness of the stationary
solution of the nonlinear hydrodynamic equation. Conversely,
it can be easily verified that it admits several solutions.
This notwithstanding, we have found only one solution that does not
yield unphysical divergences in the
long time and large space limits (see Section III). 
Accordingly, we have assumed
that fluctuations can be meaningfully estimated only with respect to
this solution. Specifically, our statistical non-equilibrium measure
is constructed by considering ``trajectories'' of the transverse
component of the velocity field $ v_T(t,\vec x)$
that are connected to $ {\bar v}_T(t,\vec x)$  by the fluctuation
field $ u_T(t,\vec x) $. Let us point out that, consistently with
the linearization procedure, the statistical measure is assumed to
be concentrated around $ {\bar v}_T(t,\vec x)$.
One further technical problem is that the linearized hydrodynamic
operator has coefficients depending on both space and time variables
through $ {\bar v}_T(t,\vec x)$, as a consequence of its originary
nonlinear nature. It depends also on the Reynolds number ${\cal R}$
in such a way that the action integral can be naturally solved by
a perturbative expansion in ${\cal R}^{-1}$. We want to remark that
sufficiently large values of ${\cal R}$ also yield a weak dependence
on time of $ {\bar v}_T(t,\vec x)$. These points are discussed in Section
IV.

Since our main purpose here is the estimation of the structure function
(see Section V)
as an average over the non-equilibrium measure induced by the action
$I$, we have to assume that the perturbative expansion applies in
a wide range of values of ${\cal R}$. In particular, we guess that it holds
also for moderately large ${\cal R}$, since a statistical average of any
observable cannot be valid for too large values of ${\cal R}$, i.e. in a regime
of fully developed turbulence. We will argue that
statistical estimations can be consistently obtained for values of ${\cal R}$
which extend up to the region of stability of the solution
$ {\bar v}_T(t,\vec x) $. Beyond this region we have no practical way
of controlling the convergence of the linearization procedure.
It is worth stressing that we obtain an analytic expression of
the structure function: the so--called K41 scaling law \cite{K41} 
is recovered on a spatial
scale, whose nontrivial dependence on ${\cal R}$ is explicitly
indicated.

\section{The model}
We consider the Navier-Stokes equation for 
the velocity vector-field components $v^{\alpha}$
describing a divergence-free homogeneous isotropic flow:
\ba
\label{NS1}
&&\left({\partial\over
\partial t} - \nu\nabla^2\right)v^{\alpha}(\vec{x}, t) + v^{\beta}(\vec{x},
t){\partial\over \partial x^{\beta}}v^{\alpha}(\vec{x}, t) + {1\over
\rho}{\partial\over \partial x^{\alpha}}P(\vec{x}, t)  -
f^{\alpha}(\vec{x}, t) = 0,\\
&&\label{cons1}
{\partial\over \partial x^{\alpha}}v^{\alpha}(\vec{x}, t) = 0.
\ea
Here, the field $f^{\alpha}$ represents a source/sink of momentum necessary
to maintain velocity fluctuations. Customarily \cite{Niko98}, we assume
$f^{\alpha}$ to be a white-in-time zero-mean Gaussian random force with
covariance
\be
\label{ff}
\langle f^{\alpha}(\vec{x}, t)f^{\beta}(\vec{x}^{\prime}, t^{\prime})\rangle
= F^{\alpha\beta}\left(\vec{x} - \vec{x}^{\prime}\right)\delta\left(t -
t^{\prime}\right)\ .
\ee
A standard choice for $F$ is
\be
F(\vec{x}) ={D_0L^3\over (2\pi)^3}\int d^3p\ e^{i\vec{p}\cdot\vec{x}}(Lp)^s
e^{-(Lp)^2} ,
\label{misura}
\ee
where $D_0$ is the power dissipated by the unitary mass, $L$ is the
integral scale and  the exponent $s$ is an integer number of
order one (a typical value is $s=2$).  Due to constraint  
(\ref{cons1}), the
field
$v^{\alpha}$ depends only on the  transverse degrees of freedom of
$f^{\alpha}$. Without prejudice
of generality we can also assume divergence-free forcing yielding the
additional relation
\be
{\partial\over \partial x^{\alpha}}F^{\alpha\beta}\left(\vec{x} -
\vec{x}^{\prime}\right)
= {\partial\over \partial x^{\beta}}F^{\alpha\beta}\left(\vec{x} -
\vec{x}^{\prime}\right) = 0\ .
\ee

Following the Martin-Siggia-Rose formalism \cite{MSR} we introduce
the Navier-Stokes density of Lagrangian
\ba
{\cal L}(v, w, P, Q, f) &&= w^{\alpha}(\vec{x}, t)\left[
\left({\partial\over
\partial t} - \nu\nabla^2\right)v^{\alpha}(\vec{x}, t) + v^{\beta}(\vec{x},
t){\partial\over \partial x^{\beta}}v^{\alpha}(\vec{x}, t)\right.\nonumber\\
&&\left.+ {1\over \rho}{\partial\over \partial x^{\alpha}}P(\vec{x}, t)
- f^{\alpha}(\vec{x}, t)\right] + {1\over \rho}Q(\vec{x}, t)
{\partial\over \partial x^{\alpha}}v^{\alpha}(\vec{x}, t)\ ,
\ea
where the field $w^{\alpha}$ is the conjugate variable to the velocity field
$v^{\alpha}$ and the field $Q$ is the Lagrangian multiplier related to
constraint (\ref{cons1}). In a similar way the pressure field $P$
acts as a Lagrangian multiplier of the solenoidal constraint for the
auxiliary field $w^{\alpha}$. The generating functional is given by
the integral
\ba
{\cal W}\left(P, Q, J\right)  
&&= \int {\cal D}v{\cal D}w{\cal D}f
\exp\left\{i\int dt\ d^3x\left[{\cal L}(v, w, P, Q, f) +
J_{\alpha}v^{\alpha}\right]\right. \nonumber\\
&&\left.-{1\over 2}\int dt
d^3xd^3yf^{\alpha}F^{-1}_{\alpha\beta}f^{\beta}\right\}
\ea
where $J_{\alpha}$ is an external source.
By performing successive integrations over the statistical
measure, ${\cal D}f e^{-\int fF^{-1}f}$, on the auxiliary field $Q$,
on the longitudinal
components of $w^{\alpha}$ and on the pressure field, all the
longitudinal degrees of freedom can be eliminated and we end up with the
effective functional for the transverse components (here denoted by 
$v_{T\alpha}$
and $w_{T\alpha}$)
\ba
{\cal W}(J)
&&= \int {\cal D}v_T{\cal D}w_T
\exp\left\{i\int dt\ d^3x\left[w^{\alpha}_T\left({\partial\over \partial 
t} -
\nu\nabla^2\right)v_{T\alpha}\right.\right.\nonumber\\
&&\left.\left.+ w^{\alpha}_Tv^{\beta}_T\partial_{\beta}v_{T\alpha}
+ J_{\alpha}v^{\alpha}_T\right](t, \vec{x})
-\int dt\ d^3xd^3y w^{\alpha}_T(t, \vec{x})
F_{\alpha\beta}(|\vec{x}-\vec{y}|) w^{\beta}_T(t, \vec{y})\right\}\ .
\ea
Diagramatic strategies are usually applied at this level. We want to point
out that a completely different point of view can be followed by observing
that also the transverse components of the auxiliary field
$w^{\alpha}_T$ can be integrated out and one finally obtains:
\be
{\cal W}(J) = \int {\cal D}v_T e^{-{1\over 2}I(v_T) + i\int dt d^3x
J_{\alpha}v^{\alpha}_T}
\label{funct1}
\ee
where the action functional $I$ is given by
\ba
I(v_T) &&= \int dt d^3x d^3y\left[\left({\partial\over \partial t} -
\nu\nabla^2\right)v_T^{\alpha}(t, \vec{x}) + v_T^{\rho}(t, \vec{x})
\partial_{\rho}v_T^{\alpha}(t, \vec{x})\right]\nonumber\\
&&{1\over F^{\alpha\beta}(|\vec{x}-\vec{y}|)}
\left[\left({\partial\over \partial t} -
\nu\nabla^2\right)v_T^{\beta}(t, \vec{y}) + v_T^{\lambda}(t, \vec{y})
\partial_{\lambda}v_T^{ \beta}(t, \vec{y})\right]\ .
\ea
This expression links the functional representation of the NSE
to the large deviation theory developed in
\cite{KL,BDGJL}). In particular the entropy is related to the functional
$I(v_T)$ by
\cite{BDGJL}
\be
S(v_T) = \frac{1}{2}\inf_{\bar{v}}I(\bar{v} )\ .
\ee
Where the minimum is taken over all trajectories connecting a steady-state
at time $t=-\infty$ with  the velocity field
$v^{\alpha}_T(0)$.

\section{A quasi-steady solution and its stability}
We consider the equation of the extremal condition for the functional ${\cal
W}(J)$ at $J^{\alpha} = 0$. In terms of the functional $I(v_T)$ it reads:
\ba
{\delta I(v_T)\over  \delta v^{\sigma}_T(t, \vec{x})} &&= 2\int d^3y\left[
-\delta_{\sigma}^{\ \alpha}\left({\partial\over \partial t} +
\nu\nabla^2\right) + \partial_{\sigma}v_T^{\alpha}(t,
\vec{x})\right.\nonumber\\
&&-\delta_{\sigma}^{\ \alpha}v_T^{\rho}(t,
\vec{x})\partial_{\rho}\bigg]\left({1\over 
F^{\alpha\beta}(|\vec{x}-\vec{y}|)}
\left[\left({\partial\over \partial t} -
\nu\nabla^2\right)v_T^{\beta}(t, \vec{y})\right.\right.\nonumber\\
&&+ v_T^{\lambda}(t, \vec{y})
\partial_{\lambda}v_T^{ \beta}(t, \vec{y})\bigg]\bigg) = 0\ .
\label{estremo}
\ea
We observe that every solution of the equation
\be
\left({\partial\over \partial t} -
\nu\nabla^2\right)v_T^{\beta}(t, \vec{x}) + v_T^{\lambda}(t, \vec{x})
\partial_{\lambda}v_T^{ \beta}(t, \vec{x}) =
\partial^{\beta}\Phi(t, \vec{x})\ ,
\label{ridotta}
\ee
is solution of (\ref{estremo}) as well. This is because of the presence
of the projector on the transverse
degrees of freedom in $F^{\alpha\beta}(|\vec{x}-\vec{y}|)$.
In particular the solutions of equation
(\ref{ridotta}) are forcing independent. Indeed this equation is the 
usual unforced NSE with
$\Phi\equiv {1\over\rho}P$.
We
consider now the particular case where $\partial^{\beta}\Phi = 0$, that
corresponds to search for transverse solutions satisfying the relation
\be
\partial_{\beta}\left(v_T^{\lambda}(t, \vec{x})
\partial_{\lambda}v_T^{ \beta}(t, \vec{x})\right) = 0\ .
\label{cond2}
\ee

In order to investigate the  statistics of fluctuations, the determination
of the hydrodynamic trajectory minimizing the entropy functional
has to be performed explicitly.
Unfortunately, existence and uniqueness theorems for the solution
of  eq.~(\ref{ridotta}) are not available  and
the criterion for chosing a suitable
solution can rely only upon physical considerations. Actually, we have found
several different solutions: among them, the only one unaffected by divergences
in space and time is the following:
\be
\begin{array}{ll}
V^{\alpha}(t, \vec{x}) = {U^{\alpha}\over 2}\left\{1 + e^{-{U^{2}\over
4\nu{\cal R}^2}t}\sin\left({2U\over
\sqrt{b^2U^2-(\vec{U}\cdot\vec{b})^2}}
{\left(\vec{b}\wedge\vec{U}\right)\cdot\vec{x}\over 4\nu{\cal
R}}\right)\right\}\ ,& with \quad t>0\ ,\\
V^{\alpha}(t, \vec{x}) = {U^{\alpha}\over 2}\ ,& with\quad t<0\ .
\end{array}
\label{soluz}
\ee
Notice that this solution decays exponentially in time with a rate $\nu/L^2
= \tau_D^{-1}$, where $\tau_D$ is the diffusion time scale. Moreover,
for $ |x| \ll L$ this
solution approximates a linear shear flow. This is well known to
produce small scale instabilities for sufficiently large Reynolds number.

Due to the invariance under Galileo transformations,
solution (\ref{soluz}) is determined up to two
vector parameters; the velocity
$U^{\alpha}$ and the rotation axis $b^{\alpha}$.
These constants are related to the energy
and to the enstrophy at the time $t$. Condition
(\ref{cond2}) is trivially satisfied, indeed
\be
V^{\beta}(t, \vec{x})\partial_{\beta}V^{\alpha}(t, \vec{x}) = 0\ .
\ee
In other words, eq.(\ref{soluz}) is also solution of the diffusion
equation $\left(\partial_t-\nu\nabla^2\right)u^{\alpha}(t, \vec{x}) = 0$.
Obviously (\ref{soluz}) is not a steady solution. It actually decays with a
typical time $\tau = {4\nu {\cal R}^2\over U^2}$ which increases with
${\cal R}$~. We assume to consider sufficiently large values of the
Reynolds number (or, equivalently, sufficiently small values of $U^2$)
so that (\ref{soluz}) can be viewed as a quasi-steady
solution.

Small fluctuations around this quasi-steady solution can be analyzed by
introducing the fluctuation field $u^{\alpha}(t, \vec x)$:
\be
v^{\alpha}(t, \vec x) =
V^{\alpha}(t, \vec x) + u^{\alpha}(t, \vec x)
\ee
and by studying the linearized equation
\be
{\partial\over \partial t}u^{\alpha}(t,\vec{x})
- \nu\nabla^2u^{\alpha}(t,\vec{x}) +
V^{\beta}(t,\vec{x}){\partial\over \partial 
x^{\beta}}u^{\alpha}(t,\vec{x}) 
+ u^{\beta}(t,\vec{x}){\partial\over \partial x^{\beta}}
V^{\alpha}(t,\vec{x})= 0\ .
\label{2-5}
\ee
For sufficiently small initial perturbations, linear analysis
can be applied up to a time scale where the nonlinear terms
are kept small with respect to the linear ones.
An upper bound for this time scale is
computed in Appendix A:
\be
0<t\lesssim {8\nu{\cal R}\over U^2},
\label{cond1a}
\ee
This upper bound turns out to be consistently smaller than $\tau_D$.
Moreover, a sufficient condition for the stability of solution
(\ref{soluz}) with respect to small perturbations
has been also derived in Appendix A:
\be
{8\nu^2{\cal R}\over U^2}k^2> 1\ .
\label{cond1}
\ee
This condition indicates that for high Reynolds numbers
only large wave-numbers are stable.

\section{Perturbative analysis of the generating functional}
By exploiting the translational invariance of the
functional measure, (\ref{funct1}) can be rewritten in the form
\be
{\cal W}(J) = \int{\cal D}u_T\ e^{-{1\over 2}I_V(u) +i\int
dtd^3x J_{\alpha} u^{\alpha}_T} .
\label{AF1}
\ee
We recall that we aim at obtaining explicit expresions for 
the structure functions of the perturbation field around the quasi-steady
solution by performing derivatives of the functional generator
with respect to the currents $J^{\alpha}$. \\
Without any further approximation, such a program seems to be prohibitive.
Some simplifications have to be introduced, yielding a structure of the
functional integral which involves Gaussian integrations.
\\ As a first step in this direction we replace the original action
in the functional (\ref{AF1}) by the  bilinear action
\ba
I_V(u) &&= \int 
dtd^3xd^3y\left[(\partial_t-\nu\nabla^2_x)u^{\alpha}_T(\hat{x})
+V^{\rho}(\hat{x})\partial_{\rho}u^{\alpha}_T(\hat{x})
+ u^{\rho}_T(\hat{x})\partial_{\rho}V^{\alpha}(\hat{x})\right]
{1\over F^{\alpha\beta}(|\vec{x}-\vec{y}|)}\times\nonumber\\
&&\left[(\partial_t-\nu\nabla^2_y)u^{\beta}_T(\hat{y})
+V^{\lambda}(\hat{y})\partial_{\lambda}u^{\beta}_T(\hat{y})
+ u^{\lambda}_T(\hat{y})\partial_{\lambda}V^{\beta}(\hat{y})\right]
+ O(u^3)\ .
\label{FF}
\ea
which comes from the linearized equations (\ref{2-5}) and where we
have used the notation $\hat{x}\equiv (t, \vec{x})$.
Note that the extreme solution $V^{\alpha}$, around which fluctuations 
are computed, appears explicitly in the functional $I_V(u)$.
We have already observed that the solution
$V^{\alpha}$ is not unique. Nevertheless, if the stability conditions
of this solution with respect to small perturbations are fulfilled
we can conclude that it cannot be influenced by other possible solutions.\\

One further simplification can be introduced  by observing that solution
(\ref{soluz}) naturally suggests a perturbative expansion in
integer powers of ${1\over {\cal R}}$.  Indeed,
the Fourier transform of action (\ref{FF}) 
up to the first order in the expansion parameter $1\over {\cal R}$,
becomes
\be
I_V(u) = \int {d^4p\over (2\pi)^4}u^{\rho}(-\hat{p})M_{\rho}^{\ 
\alpha}(-\hat{p})
{1\over F^{\alpha\beta}(p)}M^{\beta}_{\ \zeta}(\hat{p})
u^{\zeta}(\hat{p}) + O\left({1\over {\cal R}^2}\right)\ ,
\label{act1}
\ee
where the hydrodynamic evolution term
$M^{\beta}_{\ \zeta}(\hat{p})u^{\zeta}(\hat{p})$ has the form
\be
M^{\beta}_{\ \zeta}(\hat{p})u^{\zeta}(\hat{p})
= \left\{\delta^{\beta}_{\ \zeta}\left[i\left(p_0 + {1\over 
2}\vec{p}\cdot \vec{U}\right) +\nu
p^2 -{C\over 4}\vec{p}\cdot\vec{U} 
{\left(\vec{b}\wedge\vec{U}\right)^{\gamma}\over 4\nu{\cal
R}}\partial_{p_{\gamma}}\right]  -{C\over
4}U^{\beta}{\left(\vec{b}\wedge\vec{U}\right)_{\zeta}\over
4\nu{\cal R}}\right\}u^{\zeta}(\hat{p}) .
\ee
Here $C = {2U\over \sqrt{b^2U^2-(\vec{U}\cdot\vec{b})^2}}$,
$\hat{p}\equiv (p_0,\vec{p})$ where $p_0$ and $\vec{p}$ are
the conjugate variables of $t$ and $\vec{x}$, respectively.

In order to evaluate the functional integral in (\ref{AF1}),
the diagonalization of the matrix
$M_{\rho}^{\ \alpha}(-\hat{p}){1\over F^{\alpha\beta}(p)}M^{\beta}_{\
\zeta}(\hat{p})$ is required. 
Since by definition the factor $[F^{\alpha\beta}(p)]^{-1}$ is 
proportional to the identity operator in the space of the transverse 
solutions it remains to diagonalize only the matrix 
$M^{\beta}_{\zeta}(\hat{p})$.\\
The computation  of the eigenvalues, $\lambda$, of
$M^{\beta}_{\zeta}(\hat{p})$ can be accomplished by a
standard procedure, which, however requires lenghty and
tedious calculations: they are sketched in Appendix B.  
We report hereafter the final form taken by the generating
functional:
\be
{\cal W}(\eta) = \int {\cal J}(H){\cal D}\phi_T\ e^{-{1\over 
2}\int_{\hat{p}}\phi_T^{\rho}(-\hat{p})
{\lambda^{*(\rho)}(\hat{p})\lambda^{(\rho)}(\hat{p})\over 
F(p)}\phi_T^{\rho}(\hat{p}) +i\int_{\hat{p}}
\eta_{T\alpha}(-\hat{p})\phi_T^{\rho}(\hat{p})}\
\ee
Here $H$ is the matrix that diagonalizes
$M_{\rho}^{\ \alpha}(-\hat{p}){1\over
F^{\alpha\beta}(p)}M^{\beta}_{\zeta}(\hat{p})$;
${\cal J}(H)$  is the Jacobian of the basis transformation
$u\longrightarrow \phi$, $J\longrightarrow \eta$ engendered by
$H$.

By performing a  Gaussian integration we obtain  the
{\it normalized} functional in term of the $\eta^{\alpha}$ source
\be
{\cal W}(\eta) = e^{-{1\over 
2}\int_{\hat{p}}\eta_{T\rho}(-\hat{p}){F(p)\over
\lambda^{*(\rho)}(\hat{p})\lambda^{(\rho)}(\hat{p})}\eta_{T\rho}(\hat{p})}\ 
,
\label{FJ1}
\ee
By returning to the representation in the original basis this equation can 
be rewritten as:
\be
{\cal W}(J) = e^{-{1\over 2}\int_{\hat{p}}J^{\rho}_T(-\hat{p})
\left(H{F\over 
\lambda^*\lambda}H^T\right)_{\rho\sigma}(\hat{p})J^{\sigma}_T(\hat{p})}.
\label{FJ2}
\ee
The functional (\ref{FJ2}) is the starting
point for the calculation of all correlation
functions (and structure functions), that can be
obtained by derivation with respect
to the $J^{\sigma}_T$ currents. The procedure to achieve this goal
is the subject of the next Section.

\section{Short-distance behavior of the
second order structure function}
The expression derived for the generating functional (\ref{FJ2}),
contains all the statistical information on the  fluctuations
around the basic solution $V_{\alpha}$. Here we will perform
analytic calculations for the particular class of fluctuations
captured by the lowest, nontrivial, integer moment of velocity
differences between points separated by a distance $r$.
For the velocity field, $u^{\alpha}$, this is the second-order
structure function defined as
\ba
S_2&&=\langle\left|u(t, \vec{r}+\vec{x})-u(t,\vec{x})\right|^2
\rangle\nonumber\\
&&= \langle\left|(u^{\alpha}(t,
\vec{r}+\vec{x})-u^{\alpha}(t, \vec{x}))(u_{\alpha}(t,
\vec{r}+\vec{x})-u_{\alpha}(t, \vec{x}))\right|\rangle ,
\label{struttura}
\ea
where the brackets denote averages on the forcing statistics.\\
By assuming isotropy and homogeneity of the velocity field,
expression (\ref{struttura}) is expected to assume the typical
form of scale invariant functions
\begin{equation}
S_2(r) =  r^{\zeta_2}F_2\left(t, {r\over L}\right) 
\end{equation}
Here $r = |\vec{r}|$ and  $L$ is the large spatial scale associated with the
noise source. It is worth stressing that, at variance with
fully developed turbulent regimes, here the assumption of
isotropy and homogeinity have to be taken as a plausible
hypothesis allowing for analytic computations.\\

We want to point out that any exponent $\zeta_n$ should be
independent of the basis chosen for representing
the functional ${\cal W}$. Making use of (\ref{FJ1}), one
obtains:
\ba
S_2(r) &&=
\langle\left|\vec{u}(\vec{x}+\vec{r})- \vec{u}(\vec{x}) 
\right|^2\rangle\nonumber\\
&&=\left.\left({\delta\over i\delta\eta^{\alpha}(t,\vec{x}+\vec{r})}
- {\delta\over i\delta\eta^{\alpha}(t,\vec{x})}\right)
\left({\delta\over i\delta\eta_{\alpha}(t,\vec{x}+\vec{r})}
- {\delta\over i\delta\eta_{\alpha}(t,\vec{x})}\right){\cal
W}(\eta)\right|_{\eta=0}\nonumber\\ &&=2\left(\Delta^{\alpha}_{\,\,\,
\alpha}(0,\vec{r})  - \Delta^{\alpha}_{\,\,\, \alpha}(\hat{0})\right)
= 2\int {dp_0d^3p\over (2\pi)^4}\left(e^{i\vec{p}\cdot\vec{r}}
- 1\right)\left(\Delta_{11}(\hat{p}) + 
\Delta_{22}(\hat{p})\right)\nonumber\\
&&= -2\int {dp_0d^3p\over (2\pi)^4}\left(e^{i\vec{p}\cdot\vec{r}}
- 1\right)\sum_{\alpha=1}^2{F(p)\over \left(p_0+{1\over 
2}\vec{p}\cdot\vec{U}\right)^2
+ \left(\nu p^2 +\lambda^{\alpha}_{(1)}(\vec{p},\vec{U}, \vec{b}) 
\right)^2}\ .
\label{S21}
\ea
The explicit integration over $p_0$ yields
\be
\bar{S}_2(r) = -\int {d^3p\over (2\pi)^3}{e^{i\vec{p}\cdot\vec{r}} - 1\over
\nu}\sum_{\alpha=1}^2 {F(p)\over p^2 +{1\over 
\nu}\lambda^{\alpha}_{(1)}(\vec{p},\vec{U}, \vec{b})+
...}\
\label{S22}
\ee
The expressions of the eigenvalues $\lambda^{\alpha}_{(1)}$ 
($\alpha = 1, 2$) are given in Appendix B.\\
In remains to specify the geometrical structure of the flow. For the
sake of simplicity, we assume that the vector $\vec{r}$ corresponds to
the polar axis and that the vector $\vec{b}$ is orthogonal to both
$\vec{r}$ and $\vec{U}$. It turns out that $S_2(r)$ can be expressed 
as the sum of two terms: the first one is associated with the null eigenvalue 
$\lambda_{(1)}^1$, while the second one depends on the nonzero eigenvalue
$\lambda_{(1)}^2$. Namely,
\begin{equation}
\bar{S}_2(r) =-{1\over \nu}\left(I_1(r) +I_2(r)\right) .
\end{equation}
The expression of $I_1(r)$ is derived in Appendix C:
\be
I_1(r) = D_0L^3\int{d^3p\over (2\pi)^3}\left(e^{i\vec{p}\cdot\vec{r}} - 
1\right) {(Lp)^se^{-(Lp)^2}\over
p^2}\
\ee
By simple algebraic manipulations, it can be recasted into the form
\be
I_1(r) = {D_0\over 
(2\pi)^2}r^2\sum_{n=0}^{\infty}(-1)^{n+1}{\Gamma\left({s+3+2n\over
2}\right)\over
\Gamma\left(2n+4\right)}\left({r\over L}\right)^{2n}\ .
\label{i11}
\ee
We can conclude that, for short distance $r$,
the leading contribution in
$I_1(r)$ is $r^2$, that is a dissipative contribution.\\
Some lengthy algebra (see Appendix C for details)
provides also an expression for $I_2(r)$:
\ba
I_2(r) &&= D_0L^3{32\nu^2{\cal R}\over 
U^2}\left\{\int_0^{\infty}{p^2dp\over 
(2\pi)^2}(Lp)^se^{-(Lp)^2}\int_{-1}^1dx\left(e^{iprx} -
1\right)\right.\nonumber\\
&&\left.\times\left(\sum_{l=1,2}{\left(1-x^2\right)^{1\over 3}\over 
x^{2\over 3}}
{\left[\sum_{m=0}^2s_{lm}
F_m\left(x, {8\nu^2{\cal R}\over U^2}p^2; \Sigma, \Xi\right)
 + {1\over 2}\Sigma{x^{2\over 3}\over \left(1-x^2\right)^{1\over 
3}}\right]\over
\prod_{i\not= l}\left(\sum_{k=0}^2\left(s_{lk}
-s_{ik}\right)F_k\left(x, {8\nu^2{\cal R}\over U^2}p^2; \Sigma, 
\Xi\right)\right)}
+ O\left({1\over {\cal R}^2}\right)\right)\right\} . \nonumber\\
\label{tremendo}
\ea
where the coefficients $s_{ij}$ and
the functions $F_i$ are specified in Appendix C.\\
The key remark for proceeding in this calculation is that the
stochastic measure $p^{2+s}e^{-(L p)^2}dp$ gives a significant contribution to the
first integral in (\ref{tremendo}) only in a narrow region of wavenumbers
close to $\bar{p}$ where the function $p^{2+s}e^{-(Lp)^2}$ has its maximum,
i.e.
\be
\bar{p}={1\over L}\sqrt{s+2\over 2}\ .
\ee
Notice that the function ${8\nu^2{\cal R}\over U^2}p^2$ thus contributes to the
integral by taking values close to $4(s+2)\over {\cal R}$.
Moreover, for $p=\bar{p}$ the sufficient condition (\ref{cond1}) for the 
stability of small perturbations determines the upper bound 
\be
{\cal R}\lesssim 4(s+2)\ ,
\label{stabi}
\ee
This implies that for sufficiently small Reynolds' numbers the 
wavenumber $\bar{p}$ is stable. Under this condition, the leading 
contribution in (\ref{tremendo}) can be obtained
by performing an expansion in powers of $U^2\over 8\nu^2{\cal R}p^2$.\\
One finally obtains the complete expression of the structure function
(see Appendix C for details)
\ba
\bar{S}_2(r) &&= -{1\over \nu}\left(I_1(r)+I_2(r)\right)\nonumber\\
&&\sim -{D_0\over
(2\pi)^2\nu}r^2\sum_{n=0}^{\infty}\left\{(-1)^{n+1}\Gamma\left({s+2n+3\over 
2}\right)
\left[{1+\Xi\over \Gamma\left(2n + 4\right)}
-{2^{13\over 3}\Xi\over \Sigma^{2\over 3}}{2n+4\over
\Gamma (2n+6)}\right]\left({r\over L}\right)^{2n}\right.\nonumber\\
&&\left.+ O\left({{\cal R}\over 4(2+s)}\right)\right\}\ ,\quad
\mbox{for} \quad 1<{\cal R}\ll 4(2+s)\ ,
\label{SS2}
\ea
At leading order in the distance $r$ it is dominated by a dissipative contribution.

We conjecture that this analysis can be extended
to the parameter region defined by the conditon ${\cal R}> 4(2+s)$, where 
the statistically relevant wavenumbers can be unstable.
As shown in Appendix C, in this case $I_2(r)$ gives two contributions: one
is again dissipative, while there is another one yielding the nontrivial
scaling behavior $r^{2/3}$.
Specifically, the expression of $S_2(r)$ for ${\cal R}> 4(2+s)$
is
\ba
\bar{S}_2(r)&&=-{D_0\over \pi\nu}\left\{{1+{\Xi\over 2}\over
4\pi}r^2\sum_{n=0}^{\infty}(-1)^{n+1}{\Gamma
\left({s+2n+3\over 2}\right)\over \Gamma(2n+4)}
\left({r\over L}\right)^{2n}\right.\nonumber\\
&&\left.+ {{\cal R}^{1\over 3}\over \Gamma\left({2\over 3}\right)}
\left({\nu\over U}\right)^{4\over 3}r^{2\over 3}
\sum_{n=0}^{\infty}C_n(\Sigma)\Gamma\left({3s+3n+5\over 
6}\right)\left({r\over
L}\right)^n  + O\left({4(2+s)\over {\cal R}}\right)\right\}
\label{last}
\ea
This is dominated by the term $r^{2/3}$ for sufficiently small distances. 
Indeed, 
the crossover scale between the $r^2$ and the $r^{2\over 3}$ terms
occurs at
\be
{r\over L}\sim F{\cal R}^{-{3\over 4}}\ ,
\ee
In appendix C we evaluate the constant $F\sim 0.6$ and we report also the 
expression of the numerical coefficient $C_0(\Sigma)$ (the general
expression of the coefficients $C_n(\Sigma)$ appearing in (\ref{last}), 
has been skipped, because it has no practical interest).

It is a remarkable fact that $S_2$ can exhibit the scaling 
behavior predicted by K41 theory, which is assumed to hold 
when the velocity fluctuations are turbulent
in the so-called inertial range of scales. This suggests
that hydrodynamic fluctuations in a system at the very
initial stage of instability development already contain
some properties attributed to the developed turbulence regime.

\section{Conclusions}

In this paper we have exploited field-theoretic calculations
to  reformulate the random forced Navier--Stokes
problem in terms of a quadratic action functional. At a formal level,
the latter has the same structure of the action describing thermal
fluctuations in irreversible stationary processes.
The crucial step for obtaining the hydrodynamic
evolution operator which appears in the
action functional, is the integration over all longitudinal components
of both velocity and associated auxiliary fields. With respect
to the standard formulation which is the starting point
for diagramatic strategies, we thus perform one more field
integration.
The positive definite kernel in the action functional appears in the form
of the inverse of the forcing correlation function.\\
In terms of the action functional, the knowledge of the
whole velocity statistics
reduces to the computation of functional integrals. However,
finding an explicit solution for these integrals is quite
a difficult task, due to the nonlinear character of the problem. 
This forces us to introduce some approximations. The starting point is the
identification of a stationary solution
around which we linearize the evolution operator. We define also
a fluctuation field (with respect to the  stationary solution)
and we are able to compute (perturbatively, in the inverse of the
Reynolds number) the functional integrals over such fluctuation field.

In principle, the strategy might be applied to evaluate any
velocity multipoint statistical quantity. In order to reduce
the complexity of the algebraic manipulations we limited ourselves
to the calculation of the two--point second order momentum of velocity. 
Indeed, we aim at understanding if fluctuations at the early stage of their
development (accordingly, we dub them as pre-turbulent fluctuations)
already contain some important features of developed
turbulence. We are interested, in particular, to scale invariance.
In this respect, we find that fluctuations are organized at
different scales in a self--similar way. Remarkably, the scaling exponent
coincides with the dimensional prediction of the Kolmogorov 1941
theory \cite{K41} valid for developed turbulence regimes. Whether or not such
exponent is a genuine reminescence of the developed turbulence
phenomenology needs further investigations. \\
Unfortunately, the complexity of the derivation leading to the
K41 scaling law does not
allow to identify precisely the very origin of such a dimensional
prediction. We can however argue a relationship between the
observed dimensional scaling and the conservation laws (for momentum
and energy) associated to the two eigenvalues of the matrix appearing
in the action functional (\ref{act1}).\\
Another point to be emphasized is that the pressure term
does not play any role in the derivation of the dimensional scaling law. 
This is just a consequence of the fact that all the longitudinal 
degrees of freedom can be averaged out from the very beginning of
the computation.\\
We want to conclude by outlining some open problems and perspectives.
A first question concerns the relevance to be attributed to
the solution around which we linerize the evolution
operator. On one side we do not see any rigorous mathematical
motivation for invoking the need of a unique solution. Just heuristic
arguments based on physical considerations allowed us to identify
the selected solution.
Indeed, it represents a shear-like solution, which is a well-known
generator of instability towards smaller and smaller scales.\\
Another interesting point concerns the computation of the
third-order moment of the velocity correlators. In this case the
predictions of our approach could be compared with the $4/5$-law, which
is one among the very few exact results of turbulence theories.\\
Finally, the extension of our results to other classes of transport
problems, including passive scalar advection, could provide
a better understanding of the basic mechanism at the origin of
the observed scaling behaviors.

\begin{acknowledgments}
This work has been supported by Cofin 2001, prot. 2001023848 (AM)
and  prot. 2001021158 (RC).
We acknowledge useful discussions with G. Jona-Lasinio, M. Vergassola and 
P. Constantin.
\end{acknowledgments}

\begin{appendix}
\section{}

In this Appendix we study the stability analysis of the solution
$V^{\alpha}$  relative to the
linearized equation (\ref{2-5}). We have already observed that
$V^{\alpha}$ is a quasi-steady solution for a time \
$t\ll\tau_D={4\nu{\cal R}^2\over
U^2}$.
The Fourier transform of eq.(\ref{2-5})
with respect to the $\vec{x}$ spatial variable yields:
\ba
&&{\partial\over \partial t}\tilde{u}^{\alpha}\left(t, \vec{k}\right)
- \nu k^2\tilde{u}^{\alpha}\left(t, \vec{k}\right)
+{i\over 2}\vec{k}\cdot\vec{U}\tilde{u}^{\alpha}\left(t, \vec{k}\right)
+ {1\over 4}e^{-{t\over \tau_D}}\left\{U^{\beta}k_{\beta}
\left[\tilde{u}^{\alpha}\left(t, \vec{k}
- C{\vec{b}\wedge\vec{U}\over 4\nu{\cal R}}\right)\right.\right.\nonumber\\
&&\left.\left.- \tilde{u}^{\alpha}\left(t, \vec{k}
+ C{\vec{b}\wedge\vec{U}\over 4\nu{\cal R}}\right)\right]
+ U^{\alpha}C{\left(\vec{b}\wedge\vec{U}\right)_{\beta}\over 4\nu{\cal R}}
\left[\tilde{u}^{\beta}\left(t, \vec{k} - C{\vec{b}\wedge\vec{U}\over 
4\nu{\cal R}}\right)
+ \tilde{u}^{\beta}\left(t, \vec{k}
+ C{\vec{b}\wedge\vec{U}\over 4\nu{\cal R}}\right)\right]\right\}\nonumber\\
&&= 0\ .
\ea
By performing a perturbative expansion up to second order in the parameter 
${\cal R}^{-1}$, one obtains the system of equations
\ba
\label{AA1}
&&{\partial\over \partial t}\tilde{u}^{\alpha}_{(0)}\left(t, 
\vec{k}\right) + \nu k^2\tilde{u}^{\alpha}_{(0)}\left(t,
\vec{k}\right) +{i\over 
2}\vec{k}\cdot\vec{U}\tilde{u}^{\alpha}_{(0)}\left(t, \vec{k}\right) = 
0\ ,\\
\label{AA2}
&&{\partial\over \partial t}\tilde{u}^{\alpha}_{(1)}\left(t, 
\vec{k}\right) + \nu k^2\tilde{u}^{\alpha}_{(1)}\left(t,
\vec{k}\right) +{i\over 
2}\vec{k}\cdot\vec{U}\tilde{u}^{\alpha}_{(1)}\left(t, 
\vec{k}\right)\nonumber\\
&&={1\over 
2}\vec{k}\cdot\vec{U}C{\left(\vec{b}\wedge\vec{U}\right)_{\beta}\over 
4\nu{\cal R}}
{\partial\over \partial k_{\beta}}\tilde{u}^{\alpha}_{(0)}\left(t, 
\vec{k}\right)
-{1\over 2}U^{\alpha}C{\left(\vec{b}\wedge\vec{U}\right)_{\beta}\over 
4\nu{\cal R}}
\tilde{u}^{\beta}_{(0)}\left(t, \vec{k}\right)\ ,\\
\label{AA3}
&&{\partial\over \partial t}\tilde{u}^{\alpha}_{(2)}\left(t, 
\vec{k}\right) + \nu k^2\tilde{u}^{\alpha}_{(2)}\left(t,
\vec{k}\right) +{i\over 
2}\vec{k}\cdot\vec{U}\tilde{u}^{\alpha}_{(2)}\left(t, 
\vec{k}\right)\nonumber\\
&&={1\over 
2}\vec{k}\cdot\vec{U}C{\left(\vec{b}\wedge\vec{U}\right)_{\beta}\over 
4\nu{\cal R}}
{\partial\over \partial k_{\beta}}\tilde{u}^{\alpha}_{(1)}\left(t, 
\vec{k}\right)
-{1\over 2}U^{\alpha}C{\left(\vec{b}\wedge\vec{U}\right)_{\beta}\over 
4\nu{\cal R}}
\tilde{u}^{\beta}_{(1)}\left(t, \vec{k}\right)\ ,\\
&&...................................\nonumber
\ea
This system of equations yields the perturbative solution
\ba
\tilde{u}^{\alpha}\left(t, \vec{k}\right) &&= e^{-\left(\nu k^2+{i\over
2}\vec{U}\cdot\vec{k}\right)t}\Bigg\{F^{\alpha}_{(0)}\left(\vec{k}\right)+ 
F^{\alpha}_{(1)}\left(\vec{k}\right)\nonumber\\
&&+ C{\vec{k}\cdot\vec{U}\over 8\nu{\cal 
R}}\left[\left(\vec{b}\wedge\vec{U}\right)\cdot\vec{\nabla}_k
F^{\alpha}_{(0)}\left(\vec{k}\right) t -{U^{\alpha}\over 
\vec{k}\cdot\vec{U}}
\left(\vec{b}\wedge\vec{U}\right)\cdot\vec{F}_{(0)}\left(\vec{k}\right)t\right.\nonumber\\
&&\left.- 
\left(\vec{b}\wedge\vec{U}\right)\cdot\vec{k}F^{\alpha}_{(0)}\left(\vec{k}\right)\nu 
t^2\right]
+ O\left({1\over {\cal R}^2}\right)\Bigg\}\ .
\label{sol-0}
\ea
For the expansion (\ref{sol-0}) to be meaningful,
the time $t$ must be smaller than  $\sim {\cal R}$.
This amounts to impose the condition:
\be
0<t\lesssim {8\nu{\cal R}\over U^2}=\frac{2\tau_D}{{\cal R}}   \ .
\ee
This stability condition implies that, at any time $t$, the
linear term in the curly brackets cannot
overtake the exponential factor.
Such a requirement can be traslated into the following spectral condition
\be
{8\nu^2{\cal R}\over U^2}k^2>1\ .
\ee

\section{}

In this Appendix we sketch the calculation of the eigenvalues of the
matrix
$M^{\beta}_{\zeta}(\hat{p})$. We exploit a perturbative
approach, whose
expansion parameter is ${1\over \cal R }$. The matrix
$M^{\beta}_{\zeta}(\hat{p})$
acts on the two-dimensional space of the transverse functions and on
the one-dimensional space of the longitudinal functions.
Only the transverse degrees of freedom are physically
meaningful.\\ In terms of the $1\over \cal R$ expansion we have
\be
M = M_{(0)} + M_{(1)} + ...
\ee
where
\ba
M^{\alpha}_{(0)\ \beta} &&= \delta^{\alpha}_{\ \beta}\left[i\left(p_0
+ {1\over 2}\vec{p}\cdot \vec{U}\right)
+\nu p^2 \right]\ ,\nonumber\\
M^{\alpha}_{(1)\ \beta} &&= -\delta^{\alpha}_{\ \beta}{C\over 
4}\vec{p}\cdot\vec{U}
{\left(\vec{b}\wedge\vec{U}\right)^{\gamma}\over 4\nu{\cal 
R}}\partial_{p_{\gamma}}
-{C\over 4}U^{\alpha}{\left(\vec{b}\wedge\vec{U}\right)_{\beta}\over 
4\nu{\cal R}}\ .
\ea
A complete orthonormal basis in $R^3$ is given by the vectors
\ba
\Pi_1^{\alpha} &&= {\left(\vec{b}\wedge\vec{p}\right)^{\alpha}\over 
\sqrt{f(p)}}\ ,
\nonumber\\
\Pi_2^{\alpha} &&= {g(p)\left(\vec{b}\wedge\vec{p}\right)^{\alpha}
- f(p)\left(\vec{U}\wedge\vec{p}\right)^{\alpha}\over
\sqrt{f(p)}\sqrt{f(p)h(p)-g^2(p)}},\nonumber\\
\Pi_3^{\alpha} &&= {p^{\alpha}\over p}\ ,
\ea
where we have defined
\ba
f(p) = b^2p^2-(\vec{b}\cdot\vec{p})^2, \quad g(p) = (\vec{b}\cdot\vec{U})p^2
- (\vec{b}\cdot\vec{p})(\vec{U}\cdot\vec{p}), \quad h(p) =
U^2p^2-(\vec{U}\cdot\vec{p})^2\ .
\ea
Here, $\Pi_1^{\alpha}$ and $\Pi_2^{\alpha}$ span the transverse
subspace, while
$\Pi_3^{\alpha}$ spans the longitudinal one. Likewise,
the eigenvalues can be represented in terms of a perturbative expansion
as
\be
\lambda^a = \lambda^a_{(0)} + \lambda^a_{(1)} + ...\quad where\quad a=1, 
2, 3\ .
\ee
The zero-order eigenvalues $\lambda^a_{(0)}$ are degenerate and 
have the form
\be
\lambda^a_{(0)} =\left(i\left(p_0 + {1\over 2}\vec{p}\cdot 
\vec{U}\right) +\nu
p^2\right)\ .
\ee
The evaluation of the first order corrections $\lambda^a_{(1)}$ requires the
diagonalization of the matrix with elements $M_{(1)ij}=\left(\Pi_i,
M_{(1)}\Pi_j\right)$,  ($i,j=1, 2, 3$). After some simple but lengthy
calculations we find
\ba
\lambda^1_{(1)} &&= {1\over 2}\left(M_{(1)11}+ M_{(1)22}
- \sqrt{\left(M_{(1)11}+ M_{(1)22}\right)^2 + 4M_{(1)21}M_{(1)12}}\right)\
,\nonumber\\
\lambda^2_{(1)} &&= {1\over 2}\left(M_{(1)11}+ M_{(1)22}
+ \sqrt{\left(M_{(1)11}+ M_{(1)22}\right)^2 + 4M_{(1)21}M_{(1)12}}\right)\
,\nonumber\\
\lambda^3_{(1)} &&= M_{(1)33}\ ,
\ea
and
\ba
M_{(1)11} &&={C\over 16\nu{\cal R}}{\left(\vec{b}\wedge\vec{U}\right)
\cdot\vec{p}\over f(p)}w(p)\ ,\nonumber\\
M_{(1)22} &&= - {C\over 16\nu{\cal 
R}}{\left(\left(\vec{b}\wedge\vec{U}\right)
\cdot\vec{p}\right)\left(\vec{b}\cdot\vec{p}\right)g(p)\over
f(p)\left(f(p)h(p)-g^2(p)\right)}\left[\left(\vec{p}\cdot\vec{U}\right)w(p)
+ (\vec{b}\cdot\vec{U})g(p) -U^2f(p)\right],\nonumber\\
M_{(1)12} &&=- {C\over 16\nu{\cal
R}}{\left(\left(\vec{b}\wedge\vec{U}\right)\cdot\vec{p}\right)
\left(\vec{b}\cdot\vec{p}\right)\over
f(p)\sqrt{f(p)h(p)-g^2(p)}}\left[\left(\vec{b}\cdot\vec{U}\right)g(p)
+2\left(\vec{p}\cdot\vec{U}\right)w(p)
-U^2f(p)\right],\nonumber\\
M_{(1)21} &&= -{C\over 16\nu{\cal 
R}}{\left(\left(\vec{b}\wedge\vec{U}\right)
\cdot\vec{p}\right)\over f(p)\sqrt{f(p)h(p)-g^2(p)}}
\left(\vec{b}\cdot\vec{U}\right)\left[\left(\vec{b}\cdot\vec{p}\right)g(p)
- \left(\vec{p}\cdot\vec{U}\right)f(p)\right],\nonumber\\
M_{(1)33} &&= -{C\over 16\nu{\cal R}}{\left(\vec{p}\cdot\vec{U}\right)\over
p^2}\left(\left(\vec{b}\wedge\vec{U}\right)\cdot\vec{p}\right)\ ,
\ea
where we have defined
\be
w(p) = b^2(\vec{p}\cdot\vec{U}) - 
(\vec{b}\cdot\vec{p})(\vec{b}\cdot\vec{U})\ .
\ee
With the particular choice performed in Section V, the two physically 
relevant eigenvalues are
\ba
\lambda^1_{(1)} &&= 0\ ,\nonumber\\
\lambda^2_{(1)} &&= {U^2\over 16\nu {\cal R}}\left\{
\sin\theta_U\cos\theta_U\left[\cos^2\phi_U + 
\cos\left(2(\phi_U-\phi)\right)\right]
\sin^2\theta
\right.\nonumber\\
&&\left.+\cos^2\theta_U\sin 2\theta\cos(\phi_U -\phi)\right\}\ .
\label{B1}
\ea

\section{}

This Appendix contains the essential steps necessary
for computing the second order
structure functions.\\ We start from the expression of $\bar{S}_2$
\ba
\bar{S}_2(r) &&= -\int {d^3p\over (2\pi)^3}{e^{i\vec{p}\cdot\vec{r}} - 
1\over
\nu}\sum_{\alpha=1}^2 {F(p)\over p^2 +{1\over \nu}\lambda^{\alpha}_{(1)}(
\vec{p},\vec{U}, \vec{b})}\nonumber\\
&&=-{1\over \nu}\left(I_1(r) +I_2(r)\right)\ ,
\label{C1}
\ea
By considering the explicit expressions of the statistical function $F(p)$ and 
of the eigenvalues $\lambda^{\alpha}_{(1)}$ (see eq.(\ref{B1})~), one has
\ba
\label{C2}
I_1(r) &&= D_0L^3\int{d^3p\over (2\pi)^3}\left(e^{i\vec{p}\cdot\vec{r}}
- 1\right) {(Lp)^se^{-(Lp)^2}\over p^2}\ , \\
I_2(r) &&= D_0L^3\int{d^3p\over (2\pi)^3}
{\left(e^{i\vec{p}\cdot\vec{r}}
- 1\right)(Lp)^se^{-(Lp)^2}\over p^2+ {U^2\over 16\nu^2 {\cal R}}
\left[2\sin\theta_U\cos\theta_U\sin^2\theta\cos^2\phi
+\cos^2\theta_U\sin 2\theta\cos \phi\right]}\ .\nonumber\\
\label{C3}
\ea
By exploiting translational invariance,  in (\ref{C3}) we have applied the 
transformation $\phi_U-\phi\rightarrow
-\phi$. The evaluation of (\ref{C2}) follows from a standard
procedure:
\ba
I_1(r) &&= D_0L^3\int{d^3p\over (2\pi)^3}\left(e^{i\vec{p}\cdot\vec{r}}
- 1\right) {(Lp)^se^{-(Lp)^2}\over p^2}\nonumber\\
&&= {D_0L^2\over 2\pi^2}\sum_{n=1}^{\infty}{(-1)^n\over
(2n)!(2n+1)}\left({r\over L}\right)^{2n}\int_0^{\infty} d\zeta\
\zeta^{s+2n}e^{-\zeta^2}\nonumber\\
&&= {D_0\over (2\pi)^2}r^2\sum_{n=0}^{\infty}(-1)^{n+1}
{\Gamma\left({s+3+2n\over 2}\right)\over
\Gamma\left(2n+4\right)}\left({r\over L}\right)^{2n}\ .
\ea
Concerning the term $I_2(r)$ we first perform the integration in the 
$\phi$ variable.
Namely,
\be
I_2(r) = D_0L^3\int_0^{\infty}{p^2dp\over (2\pi)^3}(Lp)^se^{-(Lp)^2}
\int_{-1}^{+1} d(\cos\theta)\left(e^{ipr\cos\theta} - 1\right)I_0
\label{C5}
\ee
where
\ba
I_0 &&=\int_0^{2\pi}{d\phi\over p^2+ {U^2\over 16\nu^2 {\cal R}}
\left[2\sin\theta_U\cos\theta_U\sin^2\theta\cos^2\phi
+\cos^2\theta_U\sin 2\theta\cos \phi\right]}\nonumber\\
&&=-i{32\nu^2{\cal R}\over U^2}\int_{\gamma}{zdz\over 
az^4+bz^3+cz^2+bz+a}\ ,
\label{C6}
\ea
with  $z=e^{i\phi}$ and the integration is on the unitary circle. The
coefficients $a, b, c$ are given by
\ba
&&a = \sin\theta_U\cos\theta_U\sin^2\theta\ ,\quad
b=2\cos^2\theta_U\sin\theta\cos\theta\nonumber\\
&&c = {32\nu^2{\cal R}\over U^2}p^2 +
2\sin\theta_U\cos\theta_U\sin^2\theta\ .
\label{C7}
\ea
The evaluation of the integral (\ref{C6}) requires the knowledge of
the root of a fourth degree
algebrical equation. By exploiting the Euler method \cite{EUL} we end up
with the expression
\ba
z_i&&=z_i\left(x, {8\nu^2{\cal R}\over U^2}p^2; \Sigma, 
\Xi\right)\nonumber\\
&&={x^{1\over 3}\over \left(1-x^2\right)^{1\over 6}}\left[\sum_{l=0}^2s_{il}
F_l\left(x, {8\nu^2{\cal R}\over U^2}p^2; \Sigma, \Xi\right)
+ {1\over 2}\Sigma{x^{2\over 3}\over \left(1-x^2\right)^{1\over
3}}\right]\nonumber
\quad \quad i=1, 2, 3, 4\ .
\label{C8}
\ea
The following definition has been adopted:
\ba
F_l&&=F_l\left(x, {8\nu^2{\cal R}\over U^2}p^2; \Sigma, 
\Xi\right)\nonumber\\
&&=
\left\{{\Sigma^{2\over 3}\over 12}\left[ {81\over 4}\Sigma^4 {x^4\over
\left(1-x^2\right)^2} + {81\over 2}\Sigma^2{x^2\over \left(1-x^2\right)}
 - 90\right.\right.\nonumber\\
&&-{64\over \Sigma^2}{1-x^2\over x^2} + {8\nu^2{\cal R}\over U^2}p^2
\left(189{\Sigma^2\over \Xi}{x^2\over \left(1-x^2\right)^2} + {382\over
\Xi\left(1-x^2\right)} -120{\Sigma^2\over \Xi\ x^2}\right)\nonumber\\
&&\left.+\left({8\nu^2{\cal R}\over U^2}p^2\right)^2
\left({504\over \Xi^2\left(1-x^2\right)^2} + {47\ \Sigma^2\over \Xi^2\
x^2\left(1-x^2\right)}\right) +
\left({8\nu^2{\cal R}\over U^2}p^2\right)^3{32\ \Sigma^2\over \Xi^3\ x^2
\left(1-x^2\right)^2}\right]^{1\over 3}\nonumber\\  
&&\times\left(\epsilon^l\left[1 + \left(1 -4\times 27\ h\right)^{1\over
2}\right]^{1\over 3}+\epsilon^{l-3}\left[1 - \left(1 -4\times 27\ 
h\right)^{1\over
2}\right]^{1\over 3}\right) + {1\over 2}{\Sigma^{4\over 3}x^{4\over 3}\over
\left(1-x^2\right)^{4\over 6}}\nonumber\\
&&\left.+{1\over 3}
{\left(1-x^2\right)^{1\over 3}\over \Sigma^{2\over 3}x^{2\over 3}} + 
{8\nu^2{\cal
R}\over U^2}p^2{2\over 3\Sigma^{2\over 3}\Xi\ x^{2\over 
3}\left(1-x^2\right)^{4\over
6}}\right\}^{1\over 2}\ ,
\label{C9}
\ea
with
\ba
\label{C10}
&&x=\cos\theta\ ,\quad
\Xi = \sin\theta_U\cos\theta_U\ ,\quad \Sigma =\cot\theta_U\ ,\\
\label{C11}
&&s_{il}\Leftrightarrow
\left(\begin{array}{ccc}
1 & 1 & 1\\
1 & -1 & -1\\
-1 & 1 & -1\\
-1 & -1 & -1
\end{array}
\right)\ .
\ea
Here $\epsilon$ is the cubic root of the unity: $\epsilon = {-1+ i 
\sqrt{3}\over 2}$.
Finally the function $h$ is
\ba
h &&=\left[16 +30\
\Sigma^2{x^2\over 1-x^2} + {111\over 4}\Sigma^4{x^4\over(1-x^2)^2}
+16{8\nu^2{\cal R}\over U^2}{p^2\over
\Xi}{17\Sigma^2x^2+3(1-x^2)\over (1-x^2)^2}\right.\nonumber\\
&&\left.+48\left({8\nu^2{\cal R}\over U^2}{p^2\over \Xi}\right)^2{1\over
(1-x^2)^2}\right]^3\times\left[
-128 +81\ \Sigma^4{x^4\over (1-x^2)^2} + {81\over 2}\Sigma^6{x^6\over 
(1-x^2)^3}
\right.\nonumber\\
&&-180\Sigma^2{x^2\over 1-x^2}+{8\nu^2{\cal R}\over U^2}{p^2\over \Xi}
\left(378{\Sigma^4x^4\over (1-x^2)^3}+764{\Sigma^2x^2\over
(1-x^2)^2}-240{1\over 1-x^2}\right)\nonumber\\
&&\left.+\left({8\nu^2{\cal R}\over U^2}{p^2\over \Xi}\right)^2
\left(1008{\Sigma^2x^2\over (1-x^2)^3}+94{1\over (1-x^2)^2}\right)
+64\left({8\nu^2{\cal R}\over U^2}{p^2\over \Xi}\right)^3{1\over
(1-x^2)^3}\right]^{-2}\ .
\label{C12}
\ea
Only the roots $z_1$ and $z_2$ are included into the unit circle, 
therefore
(\ref{C5}) becomes
\ba
I_2(r) &&= D_0L^3{32\nu^2{\cal R}\over U^2}\int_0^{\infty}
{p^2dp\over  (2\pi)^2}(Lp)^se^{-(Lp)^2}\int_{-1}^1dx\left(e^{iprx} -
1\right)\nonumber\\
&&\times\sum_{l=1,2}{\left(1-x^2\right)^{1\over 3}\over x^{2\over 3}}
{\left[\sum_{m=0}^2s_{lm}
F_m\left(x, {8\nu^2{\cal R}\over U^2}p^2; \Sigma, \Xi\right)
+ {1\over 2}\Sigma{x^{2\over 3}\over \left(1-x^2\right)^{1\over
3}}\right]\over
\prod_{i\not= l}\left(\sum_{k=0}^2\left(s_{lk}-s_{ik}\right)
F_k\left(x, {8\nu^2{\cal R}\over U^2}p^2; \Sigma, \Xi\right)\right)}\ .
\label{C13}
\ea
As we have already observed in Section V, only the values of the
variable $p$ around
$\bar{p}= {1\over L}\sqrt{s+2\over 2}$ give a significant contribution
to the integral in (\ref{C13}). We observe that ${8\nu^2{\cal R}\over
U^2}\bar{p}^2\rightarrow{4(s+2)\over {\cal R}}$ and the
stability condition (\ref{cond1}) imposes:
\be
1<{\cal R}<4(s+2)\ .
\label{C14}
\ee
The evaluation of the leading terms is then possible by performing an
expansion in the parameter
${U^2\over 8\nu^2{\cal R}}p^{-2}\rightarrow {{\cal R}\over
8}\zeta^{-2}$ that,
by virtue of (\ref{C14}), is smaller than unity if
$\zeta<\sqrt{s+2\over 2}$.\\
 For
$\zeta>\sqrt{s+2\over 2}$ the contribution
 to the integral rapidly vanishes. For
$1<{\cal R}\ll 4(s+2)$ we obtain
\ba
\bar{S}_2(r) &&= -{1\over \nu}\left(I_1(r)+I_2(r)\right)\nonumber\\
&&\sim -{D_0\over
(2\pi)^2\nu}r^2\sum_{n=0}^{\infty}\left\{(-1)^{n+1}\Gamma\left({s+2n+3\over 
2}\right)
\left[{1+\Xi\over \Gamma\left(2n + 4\right)}
-{2^{13\over 3}\Xi\over \Sigma^{2\over 3}}{2n+4\over
\Gamma (2n+6)}\right]\left({r\over L}\right)^{2n}\right.\nonumber\\
&&\left.+ O\left({{\cal R}\over 4(s+2)}\right)\right\}\ .
\ea
By extending the validity of our calculations to ${\cal R}>4(s+2)$, we have
${8\nu^2{\cal R}\over U^2}p^2\rightarrow{8\over {\cal R}}\zeta^2<1$ for
$\zeta<\sqrt{s+2\over 2}$. As in the previous case,
we expand (\ref{C13})
in power of the parameter ${8\over {\cal R}}\zeta^2<1$ and we obtain:
\ba
I_2(r) &&\sim \Xi D_0L^2\int_0^{\infty}{d\zeta\over  
(2\pi)^2}\zeta^se^{-\zeta^2}
\int_{-1}^1dx\left(e^{i\zeta {r\over L}x} - 1\right)
\left\{{1 +{8\over {\cal R}}\zeta^2 +...\over 2}\right.\nonumber\\
&&+ {8\over {\cal R}\Xi}\sum_{l=1,2}{\left(1-x^2\right)^{1\over 3}\over 
x^{2\over 3}}
\left({\left[\sum_{m=0}^2s_{lm}
F_m\left(x, 0; \Sigma, \Xi\right) \right]\over
\prod_{i\not= l}\left(\sum_{k=0}^2\left(s_{lk}-s_{ik}\right)F_k
\left(x, 0; \Sigma, \Xi\right)\right)}\right.\nonumber\\
&&\left.\left.+ {8\over {\cal R}}\zeta^2\left.{\partial\over \partial y}
{\left[\sum_{m=0}^2s_{lm}
F_m\left(x, y; \Sigma, \Xi\right) \right]\over
\prod_{i\not= l}\left(\sum_{k=0}^2\left(s_{lk}-s_{ik}\right)F_k
\left(x, y; \Sigma, \Xi\right)\right)}\right|_{y=0} + ...\right)\right\}\ .
\label{C16}
\ea
Two different terms, $I_2(r)=I_2^A(r)+I_2^B(r)$,
can be identified in (\ref{C16}). The
evaluation of the first is straightforward and we obtain
\ba
I_2^A(r) &&\sim {\Xi D_0L^2\over 2}\int_0^{\infty}
{d\zeta\over  (2\pi)^2}\zeta^se^{-\zeta^2}\int_{-1}^1dx\left(e^{i\zeta 
{r\over L}x} -
1\right)
\left(1 +{8\over {\cal R}}\zeta^2 + ...\right)\nonumber\\
&&= {\Xi D_0\over 2(2\pi)^2}r^2\sum_{n=0}^{\infty}{(-1)^{n+1}\over 
\Gamma(2n+4)}
\left(\Gamma\left({s+2+2n\over 2}\right)+ {8\over {\cal R}}
\Gamma\left({s+5+2n\over 2}\right)\right)\left({r\over 
L}\right)^{2n}\nonumber\\
&&\times\left(1 +{4(s+2)\over {\cal R}} + ...\right)\ .
\label{C17}
\ea
The evaluation of the second term is more cumbersome.
The leading term can be recasted in the form:
\ba
I_2^B(r) &&\sim {8D_0L^2\over {\cal R}}\int_0^{\infty}{d\zeta\over  
(2\pi)^2}\zeta^s
e^{-\zeta^2}\int_{-1}^1dx\left(e^{i\zeta {r\over L}x} -
1\right){\left(1-x^2\right)^{1\over 3}\over x^{2\over
3}}\sum_{n=0}^{\infty}A_n(\Sigma)x^{2n}\ .
\label{B2}
\ea
The coefficients $A_i$ are $\Sigma$-dependent numerical constants. The 
first two of them are given by the expressions
\ba
A_0(\Sigma) &&= {1\over 16\sqrt{3}\left(1-\sin{\pi\over 6}\right)
\cos\left({1\over 3}\tan^{-1}\sqrt{26}\right)}\ ,\nonumber\\
A_1(\Sigma) &&= -{65\sin\left({2\over 3}\tan^{-1}\sqrt{26}\right)\over
512\sqrt{26}\cos^2\left({1\over 3}\tan^{-1}\sqrt{26}\right)}\Sigma^2\ ,...
\ea 
The exact form of these coefficients is however irrelevant for our 
analysis.
After some calculations we obtain
\ba
I^B_2(r) &&= {D_0{\cal R}^{1\over 3}\over \pi\Gamma\left({2\over 3}\right)}
\left({\nu\over U}\right)^{4\over 3}r^{2\over 3}
\sum_{n=0}^{\infty}C_n(\Sigma)\Gamma\left({3s+3n+5\over 6}\right)
\left({r\over L}\right)^n\ ,
\ea
where the coefficients $C_n(\Sigma)$ are depend on the constants 
$A_i$. For $n=0$ we have
\be
C_0(\Sigma) = {54\sqrt{3}-74\over 27\sqrt{3}}A_0 + {128\over
9\sqrt{3}}A_1(\Sigma)\ .
\ee
Comparing the scaling behavior of the term $I_2^B(r)$ with the term
$I_2^A(r)$ we find
\be
r\sim \left|2\times 8.328 \sqrt{\pi}{0.0336 - 0.1127\cot^2\theta_U\over 2+
\sin\theta_U\cos\theta_U}\right|^{3\over 4} {\cal R}^{-{3\over 4}}L\ .
\ee
For the expansion  in term of $1\over {\cal R}$ to be meaningful, 
the parameter $\theta_U$ must lie in a small region around 
the value $\pi\over 2$. This implies:
$$
r\sim F{\cal R}^{-{3\over 4}}L, \quad with\quad F\sim 0.6\ .
$$
\end{appendix}

\end{document}